\documentclass{llncs}
\newif\iffull\fulltrue

\usepackage{amssymb}
\usepackage{amsmath}
\usepackage{hyperref}

\usepackage[numbers,sort&compress]{natbib}
\usepackage{todonotes}
\usepackage[capitalise]{cleveref}

\makeatletter
\let\oldsubsubsection\subsubsection
\renewcommand{\subsubsection}[1]{\oldsubsubsection{#1}~\newline \smallskip\noindent%
}
\makeatother

\title{Secure Computation and Trustless \newline Data Intermediaries in Data Spaces}
\author{
  Christoph Fabianek\inst{1}\orcidID{0009-0002-4410-8796}
  \and 
  Stephan Krenn\inst{2}\orcidID{0000-0003-2835-9093} \and 
  Thomas Lor\"unser\inst{2,3}\orcidID{0000-0002-1829-4882} \and \\
  Veronika Siska\inst{4}\orcidID{0000-0002-8057-1203}
}
\institute{
  Frequentis, Vienna, Austria\\
  \email{christoph.fabianek@frequentis.at}\\
  \and
  AIT Austrian Institute of Technology, Vienna, Austria\\
  \email{\{thomas.loruenser, stephan.krenn\}@ait.ac.at} \and
  Digital Factory Vorarlberg GmbH, Dornbirn, Austria\\
  \and
  Independent Researcher \\
  \email{veronika.siska@gmail.com}
}

\begin{document}
\maketitle
\begin{abstract}
	This paper explores the integration of advanced cryptographic techniques for secure computation in data spaces to enable secure and trusted data sharing, which is essential for the evolving data economy.
	In addition, the paper examines the role of data intermediaries, as outlined in the EU Data Governance Act, in data spaces and specifically introduces the idea of trustless intermediaries that do not have access to their users' data.
	Therefore, we exploit the introduced secure computation methods, i.e. Secure Multi-Party Computation (MPC) and Fully Homomorphic Encryption (FHE), and discuss the security benefits.
	Overall, we identify and address key challenges for integration, focusing on areas such as identity management, policy enforcement, node selection, and access control, and present solutions through real-world use cases, including air traffic management, manufacturing, and secondary data use.
	Furthermore, through the analysis of practical applications, this work proposes a comprehensive framework for the implementation and standardization of secure computing technologies in dynamic, trustless data environments, paving the way for future research and development of a secure and interoperable data ecosystem.

	\keywords{Secure Computing $\diamond$ Data Spaces $\diamond$ Data Intermediaries $\diamond$ Policy Definitions $\diamond$ Decentralized Trust}
\end{abstract}

\section{Introduction}\label{sec:intro}

Data spaces are central to enabling sovereign, interoperable, and trustworthy data-sharing, which is crucial for the emerging data economy.
Although certain techniques to support data sovereignty are inherent to data spaces, the use of modern cryptography beyond the state-of-the-art can propel the concept to the next level and unleash collaboration on sensitive data.

In this paper, we focus on privacy-enhancing technologies (PETs) for computing on encrypted data, without the need to trust any third party or particular hardware; namely multiparty computation (MPC) and fully homomorphic encryption (FHE). MPC is a distributed protocol which naturally fits the federated architecture of data spaces and could therefore be an integrated part of it. FHE, on the other hand, enables computations on encrypted data without access to the secret key and thus can also be leveraged directly between two data space participants. FHE has a smaller communication overhead than MPC, but it requires more computations on a single server than MPC.

To the best of our knowledge no comprehensive analysis nor integration concept for MPC and FHE in data spaces exist, except for our preliminary approach presented in \cite{siska_integrating_2024}, especially in support of modern collaborative use cases.

We want to stress that also alternative paradigms for secure computation exist -- including, e.g., Trusted Execution Environments (TEEs) or Federated Learning (FL) --, partially offering higher efficiency and lower bandwidth requirements than FHE or MPC.
However, the reasons for focusing on those two primitives in this paper are twofold.
Firstly, approaches like FL are tailored for specific computations to be carried out, while FHE and MPC are universal in terms of expressiveness.
Secondly, especially in the case of hardware-backed TEEs such as Intel SGX\footnote{\url{https://www.intel.de/content/www/de/de/products/docs/accelerator-engines/software-guard-extensions.html}} or ARM TrustZone\footnote{\url{https://www.arm.com/technologies/trustzone-for-cortex-m}}, additional trust not only into the cryptographic mechanisms but also into the hardware manufacturer is required, which introduces an entire additional dimension for risk assessment, especially in highly regulated domains, e.g., related to patient health data.

\subsection{Our Contribution}
In this paper we therefore systematically analyze integration challenges for multiparty computation (MPC) and fully homomorphic encryption (FHE) into data spaces, to enable seamless access to secure computation technologies for processing sensitive data in a privacy preserving manner.

Furthermore, we also analyze the potential of data intermediaries facilitating end-to-end secure data sharing and processing within data spaces, therefore addressing critical challenges associated with trust and data integrity for data escrow.
The presented approach builds on MPC and FHE techniques to ensure that neither the intermediary nor the compute nodes require trust, thereby eliminating the risk of data loss or compromise.
Thus, we singnificantly extend our previous work in \citet{siska_integrating_2024} in multiple directions by including FHE and introductin trustless intermediaries.

To holistically approach the problem, we evaluate a representative set of use cases to identify a comprehensive spectrum of challenges.
Moreover, we propose a complete approach for the integration, as well as concrete methods and technologies to solve the identified challenges, and identify gaps where further research is required.

\subsection{Paper Outline}
This paper is structured as follows.
\cref{sec:prelim} gives a short review of the concepts of data spaces, MPC and FHE.
In \cref{sec:usecases} we introduce three use cases and discuss them from a deployment perspective, extracting their key characteristics and challenges.
In \cref{sec:integration} we propose a first approach for an ubiquitous and comprehensive integration of MPC and FHE into data spaces.
Based on that, potential technical solutions and research gaps for the identified challenges are discussed in \cref{sec:solutions}.
We conclude in \cref{sec:conclusion}.

\subsection{Related Work}

Related work that considers PETs in the context of data spaces is not extensive, since the latter is relatively young as a research field.

\citet{garrido_revealing_2022} conduct a systematic review on the application of privacy-enhancing technologies (PETs) for internet-of-things (IoT) data markets, including MPC. They conclude that PETs are not frequently used in this setting, despite relevant use cases; and that there is no consensus on a general architecture, in particular regarding the usage of blockchain.

\citet{agahari_business_2021} and \cite{agahari_it_2022} offer a business perspective on MPC for data sharing, building on the business model for data marketplaces from \cite{spiekermann_data_2019}. They conduct semi-structured interviews in the privacy and security domain to study the perceived value propositions, architecture and financial models \cite{agahari_business_2021}, as well as control, trust, and perceived risks \cite{agahari_it_2022}. They find that the value of MPC is seen in increased privacy, enhanced control and reduced need for trust, but that specific data sharing risks remain since the results may still reveal sensitive information. Different deployment scenarios are also described, such as the distributed, asynchronous setup that we present via data spaces in the current paper.

\citet{muller_barriers_2022} focus on federated machine learning, with an application for the automotive industry via the project Catena-X\footnote{\url{https://catena-x.net/}}.
They explore various cryptographic techniques, such as MPC and FHE, and identify usability challenges and efficiency as the primary obstacles.
They note that these technologies are lacking in user-friendliness and specialized libraries, and currently necessitate expert knowledge for specific use cases.

Besides the limited research on MPC integration into data spaces, some work on MPC on blockchain exists, with
Secret Network\footnote{\url{https://scrt.network/}} and Partisia\footnote{\url{https://partisiablockchain.com/}} (described in \cref{sec:MPCaaS}) being the most prominent candidates.
One important difference to data space integration is the lack of a registration procedure to establish trust relationships.  
Contrary to blockchain-based solutions, the MPC node pool in data spaces is open, but nodes and their attributes are certified e.g. via verifiable credentials (VCs).
Thus, MPC groups are also not necessarily random subsets, but can be chosen by attributes.
Also, there is no need for complex broadcast protocols for arbitration, and contracts can be signed without involving a blockchain.
Payment also does not necessarily need to flow through cryptocurrencies.

There is also a blockchain-based integration of FHE for data marketplaces: \citet{serrano_peer-to-peer} describe an architecture based on smart contracts and implement a case study on an Ethereum test chain, with two participants. The resulting system is slower and includes approximation errors when compared to a simple computation without any PETs, but improves data privacy.

Furthermore, there are multiple proposals for using trusted execution environments based on blockchain for data marketplaces: Sterling is based on the private blockchain Oasis\cite{hynes_demonstration_2018}, while PDS2\cite{giaretta_pds2} uses the public Ehereum chain to provide auditability.

To the best of our knowledge, concrete integration of MPC or FHE into data spaces has not been discussed in the literature and we are the first to propose a general and comprehensive treatment.
Data spaces require a fundamentally different approach to a pure blockchain based system, and can be more flexible, scalable and energy efficient compared to permissionless systems.

\section{Preliminaries}\label{sec:prelim}

We next outline some fundamental concepts.
In particular we explain the concept of data spaces as well as the idea of data intermediaries, which are both novel data governance concepts established in the European Union.
Additionally, on the technical side we introduce two important cryptographic methods from the field of privacy enhancing technologies, i.e., secure multiparty computation and fully homomorphic encryption, which substantially matured in research over the last decade and now make its way into first commercial applications.

\subsection{Data Spaces}

A data space is ``a distributed system defined by a governance framework that enables secure and trustworthy data transactions between participants while supporting trust and data sovereignty'' \cite{dssc_glossary_2023}. The goal of data spaces is to share data and data-related services via a federated data marketplace \cite{zappa_connecting_2022}. This includes data-based services, such as storage, web servers, or algorithms operating on shared data. The latter is particularly relevant for privacy-preserving and/or distributed computing approaches that respect access and usage restrictions, such as MPC.

Data spaces were introduced in computer science as a shift from a central database to storing data at the source  \cite{franklin_databases_2005}. This new way of data management, where participants retain control over their own data, is now called data sovereignty \cite{otto2022designing}. Data sovereignty is at the heart of the European data strategy and related regulations, in particular the General Data Protection Regulation (GDPR)\footnote{\url{https://eur-lex.europa.eu/eli/reg/2016/679/oj}}, the Data Governance Act\footnote{\url{https://eur-lex.europa.eu/eli/reg/2022/868/oj}}, and the Data Act\footnote{\url{https://eur-lex.europa.eu/eli/reg/2023/2854}}. The concept is also of international interest: by now, GDPR-like regulations exist in 17 countries and even more on the federal level (e.g. New York Privacy Act\footnote{\url{https://nyassembly.gov/leg/?bn=S00365}} and the California Consumer Privacy Act\footnote{\url{https://oag.ca.gov/privacy/ccpa}}); with some (e.g. South Korea's Personal Information Protection Act\footnote{\url{https://www.law.go.kr/LSW/lsInfoP.do?lsiSeq=213857&viewCls=engLsInfoR&urlMode=engLsInfoR}}) even pre-dating GDPR.

There are many initiatives supporting data space development. The International Data Spaces Association (IDSA) provided the initial concept, including the first reference architecture, the International Data Spaces Reference Architecture Model (IDS RAM). Gaia-X is taking the concept further and considers generic data products, also including services like storage or data analytics, to enable interoperability between different infrastructures. Gaia-X also develops a trust framework: a composition of policies, rules, standards and procedures based on standardized descriptions for participants and services. These are built using W3C Verifiable Credentials: cryptographically signed digital certificates that are thus tamperproof and automatically verifiable.

The Data Spaces Business Alliance (DSBA)\footnote{\url{https://data-spaces-business-alliance.eu/}}, formed by BDVA\footnote{\url{https://bdva.eu/}}, FIWARE Foundation\footnote{\url{https://www.fiware.org/}}, Gaia-X\footnote{\url{https://gaia-x.eu/}}, and IDSA\footnote{\url{https://internationaldataspaces.org/}}, aims to harmonize these efforts by providing a common technical framework (DOME) \cite{DataSpaceBusinessAlliance2023}. The Data Spaces Support Centre (DSSC)\footnote{\url{https://dssc.eu/}} contributes with coordination efforts, including a glossary and building blocks, whereas simpl\footnote{\url{https://digital-strategy.ec.europa.eu/en/policies/simpl}} focuses on creating reusable data space software. Sector-specific projects like Catena-X in the automotive industry or Manufacturing-X for manufacturing, exemplify the application of these frameworks. Promising open-source software components for data spaces are now also available, such as the Eclipse Dataspace Components (EDC), the Gaia-X cross-federation services or the Pontus-X ecosystem. These collaborative efforts are laying the groundwork for a unified, efficient, and sovereign digital ecosystem, marking significant strides toward the realization of a comprehensive Data Economy.

\subsection{Data Intermediaries}

Data intermediaries act as important components within data ecosystems, bridging the gap between data providers and data consumers while addressing critical challenges in data processing and security. They play a crucial role in ensuring compliance with regulatory frameworks and enhancing the value extracted from data through various services. These services are essential for maintaining the integrity, usability, and accessibility of data, thereby fostering a robust data economy, and promoting innovation.

Traditionally, data intermediaries have served as brokers, aggregators, and facilitators of data transactions, primarily collecting, standardizing, and cleaning data from diverse sources before providing it to organizations for value extraction. Examples include market research firms, financial data providers, and health information exchanges, which have enabled organizations to access a broader range of data, enhance their analytics capabilities, and make more informed decisions.

The European Union’s Data Governance Act (DGA)\footnote{\url{https://eur-lex.europa.eu/eli/reg/2022/868/oj}} establishes a framework for the safe and effective sharing of data across sectors and member states. According to the DGA, data intermediaries provide services that facilitate data sharing while ensuring the protection of data subjects’ rights and interests.

\begin{figure}
	\begin{center}
		\includegraphics[width=0.5\columnwidth]{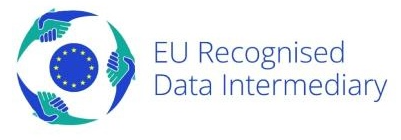}
	\end{center}
	\vspace*{-2em}
	\caption{Logo for EU Recognised Data Intermediary.}\label{fig:eu_disp}
\end{figure}

The DGA outlines a comprehensive framework for data intermediaries, specifying their roles, responsibilities, and operational conditions to ensure trustworthy data sharing. A key requirement is the mandatory notification and registration of data intermediaries with the competent national authority. To further enhance transparency and trust, the Commission has introduced a common logo for data intermediation service providers (cf. \cref{fig:eu_disp}), enabling stakeholders to easily identify compliant entities. Additionally, data intermediaries must maintain neutrality and independence, operating as neutral third parties without aggregating, enriching, or transforming data to add value. This structural separation from other services is mandated to prevent conflicts of interest and ensure that their business model does not depend on profiting directly from the data shared through them.

Article 10 of the DGA specifies three broad types of data intermediation services that can be seen as enablers of data spaces, including:
\begin{itemize}
	\item Intermediation services between data holders and potential data users, facilitating bilateral or multilateral exchanges of data;
	\item Intermediation services between data subjects or individuals and potential data users, primarily dealing with personal or non-personal data sharing; and
	\item Data cooperatives, organizational structures constituted by data subjects, one-person undertakings, or SMEs to help their members exercise their rights over their data and support collective data management and public interest goals.
\end{itemize}
These roles highlight the diverse functions of data intermediaries, from facilitating industrial data sharing to personal data management and collective data governance, underscoring their importance in the evolving data economy.

To address the challenges of data security and the risk of data loss, our approach emphasizes the use of zero-trust data intermediaries. These intermediaries leverage advanced cryptographic techniques such as MPC and FHE to facilitate secure data processing without requiring trust in the intermediary or compute nodes. This reduces the risk of data loss and enhances the integrity of the data handling process.

\subsection{Secure Multiparty Computation}

Multiparty computation (MPC) is a technology for computing on encrypted data in a distributed setting, i.e., with multiple nodes holding only secure fragments of input data not learning anything from them.
The concept appeared more than 30 years ago and has been the target of active research over the last 3 decades.
For a long time, it was considered only theoretical, but progress in recent years led to many interesting applications which can be realized with practical efficiency, given a suitable deployment.

\subsubsection{Basic Model}
In principle, MPC can be used to decentralize systems where typically a central trusted authority is needed to execute a function on behalf of the users.
With MPC, the function is evaluated jointly between multiple parties such that the correctness of the output is guaranteed and the privacy of the inputs of the individual parties is preserved; only the output of the computation is learned.
Furthermore, information-theoretically secure MPC exist which makes it the ideal method if long-term security is needed.

We quickly present the generic model of MPC as introduced in ISO/IEC 4922\footnote{\url{https://www.iso.org/standard/80508.html}}\footnote{\url{https://www.iso.org/standard/80514.html}}.
Different roles are necessary in a generic MPC system in order to qualify as such.
\textbf{Input parties} hold inputs for the secure computation which must be encoded and then sent to the compute parties.
\textbf{Compute parties} run the multiparty protocol, which is executed among them as they jointly compute the intended function on the encoded inputs.
The \textbf{intended function} to be computed is not kept secret and is defined according to the use case.
The function is composed of basic operations available to the MPC protocol and typically composed of simple gates from a boolean or arithmetic circuit, depending on the encoding and protocols used.
After the computation, the result is held by the compute parties in an encoded form and then sent to the \textbf{result parties}, which can reconstruct the result of the computation.

\begin{figure}
	\begin{center}
		\includegraphics[width=0.60\textwidth]{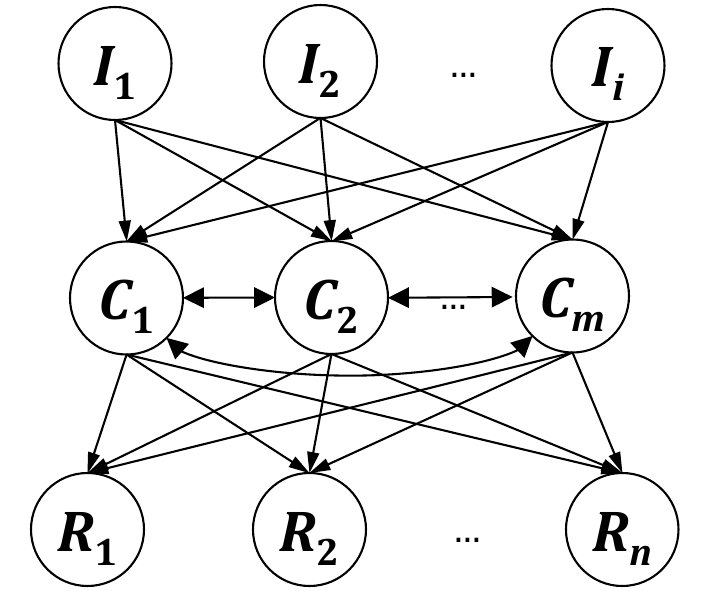}
	\end{center}
	\caption{Generic MPC model: input nodes $I_i$ encode data and send them to compute nodes $C_i$, which then execute the MPC protocol. After that, compute nodes hold the secret in encoded form, which is finally sent to result nodes $R_i$ that recover the result in plaintext, as also presented in \cite{siska_integrating_2024}.}
	\label{fig:mpcnodes}
\end{figure}

The main security properties are \textbf{correctness} and \textbf{input privacy}, and it is the latter which guarantees the confidentiality of the data.
Depending on the protocol, the security parameters could hold against different kind of adversaries.

Certain additional, optional security guarantees are also possible, e.g., fairness, guaranteed output delivery or covert security.
Fairness means, that malicious parties only receive their output if also the honest parties do so.
With guaranteed output delivery the honest parties always receive their output.
Contrary, in a covert security model, the protocol aborts in case of error and allows for cheater detection.

In summary, the overall concept is well understood and elaborated, i.e., many computations have been shown practical. However, the security assumptions are very different from traditional secret or public key cryptography. Here, security is mainly governed by the non-collusion assumption, which makes deployment of the technology challenging, especially in dynamic scenarios as we often find in emerging data markets and digital ecosystems with many stakeholders involved.

\subsubsection{MPC as a Service}\label{sec:MPCaaS}
Due to the complexity and deployment challenges, potential users are often reluctant to use MPC. Thus, collaborative use cases are often prevented in data spaces if data privacy cannot be assured.

Leveraging the as-a-service paradigm could be a way out for this problem but requires careful integration of the service to assure high security and prevent data leakage along the data life-cycle.

Moreover, additional integrity guarantees and data leakage prevention methods may be desirable depending on the sensitivity of the data and the use case.
In particular, public verifiability could be of additional value for MPC-as-a-service (MPCaaS) and contribute to the trustworthiness of the service.

Publicly verifiable MPC can assure the correctness of computations even if all compute nodes are compromised, and although input privacy does not hold anymore.
Typically, this is achieved by combining MPC protocols with compatible zero-knowledge proof (ZKP) systems to provide the best possible security guarantees for the outsourcing scenario of remote MPC, which is the case for the as-a-service usage.
Yet, this is only to prevent from corrupt results in the worst case of a fully malicious MPC system, which can be prevented by careful selection of nodes.

The possibility for public audits of computation results have additionally benefits for data spaces because it also allows for high assurance levels of computation results.
If even third-party stakeholders are able to verify the results of a computation, this could be used to establish end-to-end authenticity in data spaces. For example, \cite{kanjalkar_publicly_2021} used this concept by combining MPC and zkSNARKS \cite{DBLP:conf/eurocrypt/ChiesaHMMVW20} with universal setup to enable flexible verifiability for MPCaaS.
The idea has also been shown to be useful in the manufacturing context \cite{lorunser_privacy-preserving_2022}.


Partisia is another example which uses blockchain to persist data and as a broadcast channel in combination with an event driven architecture\footnote{\url{https://medium.com/partisia-blockchain/}}. Here, MPC node pools are built from available compute nodes, and each MPC service is randomly assigned to a subset of the nodes in the pool. Service buyers pay a pool to run a service, and the whole process is orchestrated via a smart contract, without the secret state appearing on the blockchain.

Although first proposals for MPCaaS exist, it is an open question how generic MPCaaS shall be integrated into data spaces to support a wide range of use cases, but without burdening complex configuration and deployment issues on the users of the system.
In our work we systematically analyze this problem and propose relevant technologies to be used to realize the concept.

\subsection{Fully Homomorphic Encryption}\label{sec:FHE}

Besides MPC, fully homomorphic encryption (FHE) constitutes the second main approach for cryptographically secured computation on sensitive data.

The concept of FHE was already introduced in the late 1970s by Rivest et al.~\cite{Rivest1978}, but the secure realization was proposed more than thirty years later in a groundbreaking result of Gentry~\cite{DBLP:conf/stoc/Gentry09}, followed by a large body of work focusing on key- and ciphertext sizes, efficiency, etc.
The development of FHE has seen significant advancements over the past decade, driven by improvements in both the theoretical foundations and practical implementations.
The most notable progress has been in the reduction of computational overhead, which has traditionally been a major barrier to the adoption of FHE.
Recent FHE schemes, such as those based on the BGV \cite{DBLP:conf/crypto/Brakerski12}, CKKS \cite{cheon_homomorphic_2017} and TFHE \cite{chillotti_tfhe_2020} have significantly reduced the complexity of homomorphic operations.
A range of open-source frameworks are available to researchers and developers\footnote{\url{https://fhe.org/resources/}}.

\subsubsection{Basic Model}
Fully homomorphic encryption provides significantly enhanced functionalities compared to classical encryption.
Namely, besides solely decrypting the ciphertext, it also allows one to evaluate functions in the encrypted domain.
By computing on the ciphertexts, corresponding computations on the underlying plaintext can be realized, without ever requiring access to the plaintext.
As a result, data owners can encrypt data, send it to an external (cloud) service together with a specification of the computation to be performed, and retrieve the encrypted computation result, which can be decrypted to receive the results of the computation.

In FHE, three main keys are involved: A \textbf{public key} is used to encrypt plaintext data, allowing anyone to perform encryption and homomorphic operations on ciphertexts.
The \textbf{private key} is used to decrypt the final ciphertext after homomorphic operations, ensuring that only authorized parties can access the plaintext result.
In many schemes an additional \textbf{evaluation key} facilitates efficient computation on encrypted data without needing the private key, enabling complex operations like multiplication while maintaining encryption.
Sometimes a \textbf{relinearization key} is used to manage the growth of ciphertext size during operations like multiplication. It helps keeping ciphertexts compact and secure, ensuring efficient computation without excessive increase in size.
The main security guarantees are {\bf correctness}, guaranteeing that if all entities behave honestly, the computation result will be correct, and {\bf data privacy}, guaranteeing that no information about the input data is revealed to the cloud server.

Extended guarantees like verifiability of the performed computation can be achieved by specific schemes, e.g.,~\cite{DBLP:journals/corr/abs-2301-07041}.
Furthermore, when used to analyze data coming from different data sources, also multi-key FHE schemes exist~\cite{DBLP:conf/stoc/Lopez-AltTV12}, where each participant encrypts their data under their own key;
however, despite eased key management, this approach requires all individual secret keys to be involved in the decryption process.




\subsubsection{FHE as a Service}
The concept of FHE as a Service (FHEaaS) has emerged as a promising approach to making FHE accessible to a broader range of applications and users.
Microsoft’s SEAL\footnote{\url{https://github.com/microsoft/SEAL}} and IBM’s HELib\footnote{\url{https://github.com/homenc/HElib}} are among the most prominent FHE libraries that have been integrated into cloud services.
Furthermore, startups like ZAMA\footnote{\url{https://www.zama.ai/}} and academic projects, e.g., like OpenFHE\footnote{\url{https://www.openfhe.org/}}, are also contributing to the development of FHEaaS, focusing on creating more user-friendly interfaces and improving the efficiency of homomorphic computations.

Despite these advancements, several challenges remain in the widespread adoption of FHE as a service.
The computational cost of FHE, while reduced, is still significantly higher than traditional encryption methods and will require specific hardware on the server side for high-volume or real-time applications, which is not yet available due to the lack of standardization and interoperability of FHE schemes.

Moreover, key management in FHEaaS is also a critical and complex issue, particularly in scenarios involving multiple parties and long-term data sharing.
The challenges include secure key generation, distribution, rotation, and recovery, as well as ensuring interoperability and compliance with regulatory standards.

If used to analyze data coming from different data sources, special attention needs to be paid to the management of the secret key, as it could not only be used to decrypt the computation results, but also the encrypted inputs.
Thus, in scenarios where the intended receiver of the computation result -- owning the secret key -- must not get access to the individual encrypted inputs, it needs to be ensured that the computing server does not leak encrypted input data to the receiver.
This has to be achieved on an organization level, i.e., by enforcing strict access policies and assuming non colluding servers similar to MPC.

In general, addressing key management challenges requires not only advanced cryptographic techniques but also robust infrastructure and protocols to manage keys effectively in a way that ensures both security and usability.
Threshold versions of various schemes have also been introduced and implemented \cite{OpenFHE} as well as proxy re-encryption extensions to distribute trust and relax assumptions on individual servers.
Switching between schemes is also considered a way to increase agility and flexibility in such scenarios.
Furthermore, challenges arise when exploring hybrid approaches, i.e., using FHE in conjunction MPC or other cryptographic protocols.
The keys must be managed in such a way that they enable joint computation without revealing the data to any of the participating parties.

A different approach towards outsourcing is the use of FHE on blockchain, which gained substantial attention in recent years.
As shown by Dahl et al.~\cite{fhEVM} the technology can be used to achieve various applications directly on the chain.
Among others they support encrypted tokens, blind auctions, privacy-enhancing decentralized autonomous organizations (DAOs) or decentralized identifiers (DID).
Despite the progress achieved, work ahead aims at improving the security model potentially invoking trusted hardware, support for flexible sets of validators and reduction of ciphertext size for on-chain storage.

\section{Use Cases and Challenges}\label{sec:usecases}

In the following we explore three complimentary use cares requiring secure computing technologies, and use them to identify and cluster the arising challenges.

\subsection{Use Cases}
The use cases were selected to be highly complementary, in order to derive representative challenges and requirements.

\subsubsection{Air Traffic Management}
In air traffic management, the value attributed to individual flights can significantly vary.
During peak periods, when demand exceeds available resources (e.g., due to bad weather or strikes), airlines have a vested interest in prioritizing flights that are of higher value to them.
This need aligns with the economic interests of airports, which aim for optimal utilization of their infrastructure and a steady flow of passengers.
Concurrently, air navigation service providers (ANSPs) are tasked with ensuring the safety of air travel, maintaining fairness and equality among all participants.

This scenario presents a multifaceted set of preferences and constraints, forming an optimization problem:
determining the ideal sequence of flights for arrivals and departures.
Each stakeholder -- airlines, airports, and ANSP -- has different needs, which include additional strict confidentiality requirements on which information to keep secret from other stakeholders.
In a series of works, \cite{loruenser_dasc_2021,DBLP:conf/ccsw-ws/LorunserWK22,DBLP:conf/coopis/SchuetzLJSWKG22} proposed systems to optimize the use of airport capacities while taking all stakeholders' needs into consideration.

Their approach is built on MPC to satisfy the different confidentiality and integrity needs.
In particular, verifiability of the computation is required, to minimize the risk of incorrect outputs resulting in a bias for or against a specific airline.
More generally, fairness conditions are considered, to ensure that no specific airline is systematically privileged.
Performance-wise, slot assignments are periodically computed for larger time intervals and the computation may take several minutes to succeed.

The high confidentiality and integrity needs of all stakeholders directly arise from their economic interests.
Furthermore, verifiability of the computation is required, to minimize the risk of incorrect outputs resulting in a bias for or against a specific airline.
More generally, fairness conditions are considered, to ensure that no specific airline is systematically privileged.
Performance-wise, slot assignments are periodically computed for larger time intervals and the computation may take several minutes to succeed.

The considered approaches vary slightly: \cite{DBLP:conf/ccsw-ws/LorunserWK22} output optimal solutions solving linear assignment problems, while \cite{DBLP:conf/coopis/SchuetzLJSWKG22} consider genetic algorithms that reach a near-optimal solution with high efficiency.
Independent of the precise strategy, the necessary computations are agreed upon in advance by the various stakeholders and remain fixed over a high number of executions.

On the deployment side, air traffic management turns out to be a relatively static scenario, where a steady group of input providers (i.e., airlines) contributes their preferences, and all stakeholders (e.g., compute nodes, inputs providers, output consumers, etc.) are mutually known to each other. The existence of a central trusted entity, such as the local Air Navigation Service Provider (ANSP), EUROCONTROL, or the airport itself, ensures that data integrity and confidentiality are maintained without necessitating the use of a Data Intermediary. These entities inherently manage and optimize air traffic, providing a centralized and trusted framework for the stakeholders involved. Consequently, the implementation of a Data Intermediary for secure data sharing in this sector is redundant, as the established trusted entities fulfill this role effectively.

\subsubsection{Manufacturing as a Service}
\begin{figure}
	\begin{center}
		\includegraphics[width=0.95\columnwidth,trim={0 0 90 210},clip]{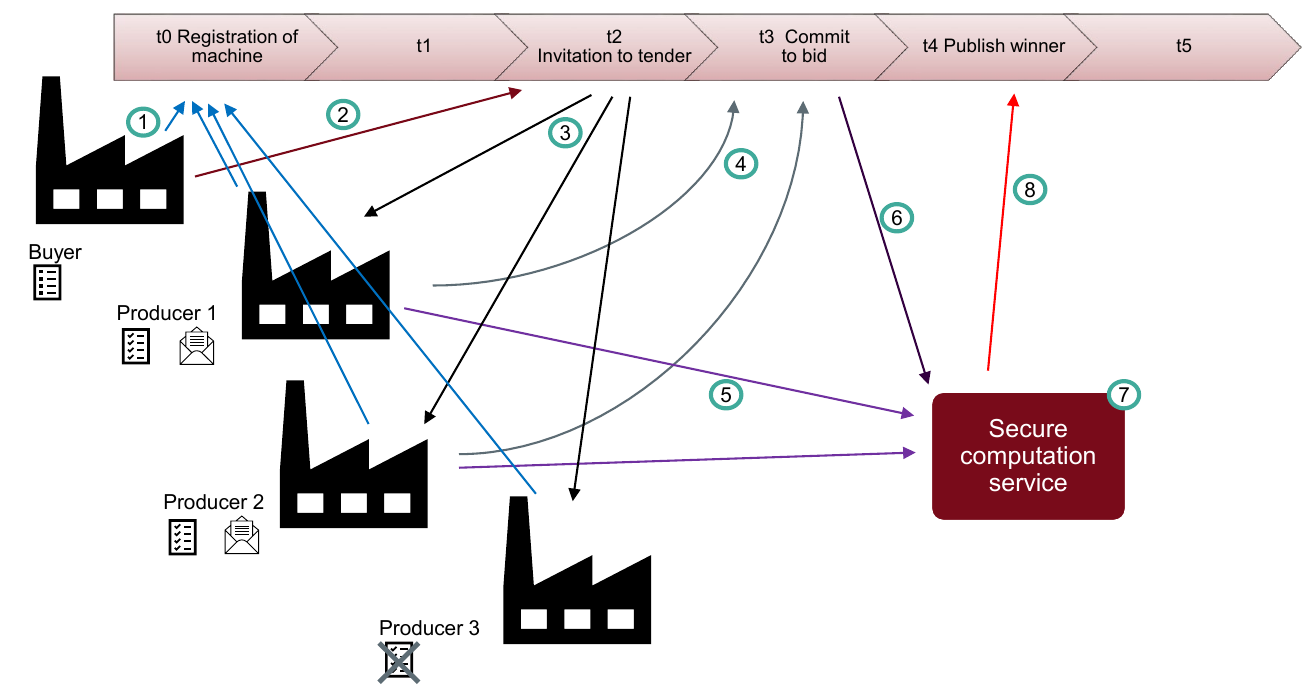} 
	\end{center}
	\caption{Manufacturing-as-a-Service Architecture adapted from \cite{lorunser_privacy-preserving_2022,siska_integrating_2024}}
	\label{fig:flexprod}
\end{figure}

The sharing economy promises environmental benefits, innovation, and reduction of costs, but concerns persist over data sovereignty and trust. Also, centralization in large infrastructures raises economic alarms.
Specifically for the manufacturing domain, \cite{lorunser_privacy-preserving_2022} examine a platform where manufacturing site owners can enlist as producers, registering their machinery along with pertinent meta information such as configurations and quality standards. Customers can place orders, prompting producers to submit bids to secure the order.
A high-level architecture and flow is depicted in \Cref{fig:flexprod}.

Regarding requirements, producers need confidentiality to make sure that non-winning bids are not leaked, to avoid exposing internal cost structures or similar information to competitors.
Both customers and producers are asking for integrity and verifiability, i.e., it needs to be ensured that the correctness of the computation can be publicly checked.
In~\cite{lorunser_privacy-preserving_2022}, this is achieved leveraging zero-knowledge proofs in combination with MPC; however, this is not to be misunderstood as a prejudice against FHE for the specific use case, but rather a specific design of the authors.
Finally, immutability of bids, to avoid adjustments depending on competing bids, is avoided using blockchain for securely storing encrypted bids and outcomes.

In the scenario of manufacturing as a service, the function to be computed is not entirely static, but may vary depending on the specific tender. For instance, while \cite{lorunser_privacy-preserving_2022} consider first-price sealed-bid auctions, also alternative options like second-price (= Vickrey) auctions or multi-attribute auctions could be used.
The precise model would be defined by the customer when publishing the tender.

From a deployment point of view, the auction platform provider would select any involved compute nodes, so that they can safely be assumed to be known a priori to all stakeholders in the default setting.
However, also more dynamic configurations could be imagined, where stakeholders which to be part of the computation to increase data sovereignty.
Moreover, as anybody may act as a customer and/or producer, the users cannot be assumed to be static and known to each other, such that a permissioned setting requiring, e.g., a registration phase, need to be introduced in order to overcome challenges with rogue bids and offers.

In this use case, where no central trusted entity exists, the role of a Data Intermediary becomes critical. Manufacturers and consumers engage in collaborative processes without a predefined trust foundation. The Data Intermediary facilitates secure data exchange using advanced cryptographic techniques such as MPC or FHE, ensuring that sensitive data is processed securely with a fixed scope, enabling trustless collaboration while safeguarding proprietary manufacturing data.

\subsubsection{Secondary Use of Data}
Data is often generated for a specific purpose, e.g., for medical treatment or collecting GPS information for charging road usage.
However, often this data would also be highly valuable in other contexts, e.g., medical studies in hospitals or road traffic planning for public authorities.
This gives raise to the concept of data marketplaces, which enable selling (computations on) data to customers.

Different approaches based on different cryptographic primitives have been proposed in the literature, e.g., using fully homomorphic encryption \cite{DBLP:journals/tdsc/KoutsosPCTH22}, or secure multiparty computation \cite{DBLP:conf/primelife/KochKPR20,DBLP:conf/de/KochKMMR22}.

According to \cite{DBLP:conf/de/KochKMMR22}, confidentiality and privacy are paramount, ensuring that (computations on) data cannot be requested without consent.
That is, data providers must have fine-grained control over data usage and sales, without relying on a single trusted entity.
Furthermore, verifiability and authenticity are crucial:
the marketplace operator should not be able to tamper with analysis outputs, and mechanisms are needed to prevent the sale of fake data to increase trustworthiness and value of data, without compromising privacy.
Where possible, end-to-end guarantees on data integrity are desirable, spanning from data source (e.g., a sensor) to consumer.

In the context of data markets, it is also crucial to support high flexibility in the computation to be carried out.
This is necessary to protect privacy and address the asynchronous nature of these ecosystems, where data providers may not be available at computation time.
Therefore, data subjects must have the ability to define precise usage policies linked to their data, specifying constraints on computations, compute nodes, and the number of inputs involved.
It is imperative that compliance with these policies is immutably documented for auditing purposes for each computation.
Additionally, contractual agreements must be in place, e.g., to prevent the acquisition of previously independent compute nodes by the same entity before data deletion.
Moreover, the trade-offs between transparency and auditability on the one hand, and customer needs on the other hand, must be carefully considered.
For instance, the mere interest of a customer in certain data may inadvertently disclose information about their business strategy.

In terms of deployment, data markets require a high level of flexibility.
Data may be stored in various locations, and users may define different types of policies, such as geographical constraints on nodes. Particularly within the health domain, a Data Intermediary is essential due to the sensitivity and legal constraints surrounding data sharing. National laws or public opinion may restrict the direct sharing of health data with supranational organizations (e.g., UN or EU). To address this, data is encoded using MPC or encrypted with FHE at the national level before being processed by a trustless intermediary. This intermediary could be setup dynamically with a pre-defined scope (e.g., addressing a rare disease) and facilitates secure and compliant data sharing while maintaining the confidentiality and sovereignty of the original data sources.

Additionally, in contrast to the previous use case domains, node selection becomes a complex task.
It is also uncertain which nodes will require access to which shares during data creation and storage, necessitating the deployment of advanced encryption mechanisms and related key management procedures to support this dynamism.
Furthermore, since data providers and consumers are typically unknown to each other, strong identity management mechanisms are essential.
These mechanisms not only ensure that users' policies (e.g., ``only medical research institutes may request computations on my data'') are adhered to, but also mitigate the risks associated with rogue data.
Finally, potential payments for data usage must be executed in a manner that preserves privacy.

\subsection{Challenges}

As illustrated by the application scenarios above, integrating MPC into complex federated scenarios such as data spaces comes with practical challenges that may directly influence system design. In the following we cluster the lessons learned from the considered use cases to obtain a set of challenge categories to be considered, which are also summarized in \cref{tab:challenges_map}.

\begin{table*}[t!]
	\centering \footnotesize
	\setlength{\arrayrulewidth}{0.5mm}
	\setlength{\tabcolsep}{3pt}
	\renewcommand{\arraystretch}{1.33}
	\begin{tabular}{|p{3.3cm}|p{2.7cm}|p{2.7cm}|p{2.7cm}|}
		\cline{2-4}
		\multicolumn{1}{c|}{}                          & \multicolumn{1}{p{2.7cm}|}{\bf UC1: \newline Air traffic}                  & \multicolumn{1}{p{2.7cm}|}{\bf UC2: \newline Industry 4.0} & \multicolumn{1}{p{2.7cm}|}{\bf UC3: \newline Secondary Use} \\
		\hline
		\bf C1. Global system parameters               & \multicolumn{2}{p{5.4cm}|}{CRS for end-to-end verifiability and integrity} &                                                                                                                          \\
		\hline
		\bf C2. Authentication and identity management & static, permissioned                                                       & semi-static, permissioned                                  & dynamic, permissionless                                     \\
		\hline
		\bf C3. Data usage policies                    & \multicolumn{2}{p{5.4cm}|}{static, fully defined from beginning}           & dynamic,  meta-level specifications                                                                                      \\
		\hline
		\bf C4. Node selection                         & static                                                                     & \multicolumn{2}{p{5.4cm}|}{dynamic}                                                                                      \\
		\hline
		\bf C5. Access control                         & online input provisioning; early encoding                                  & synchronous input; early encoding; audit info              & asynchronous input; late encoding                           \\
		\hline
		\bf C6. Trustless intermediaries               & not relevant                                                               & fixed scope                                                & dynamic deployment                                          \\
		\hline
	\end{tabular}
	\caption{Comparison of challenges affected by different use cases.}
	\label{tab:challenges_map}
\end{table*}

\paragraph{C1. Global system parameters.}
In case that the protocols to be executed require global system parameters -- such as a common reference string (CRS) -- the security and trustworthiness of these parameters needs to be guaranteed.
This may for instance apply when leveraging zkSNARKs to obtain public verifiability of the computation output.

\paragraph{C2. Authentication and identity management.}
Identity management is at the core of any security architecture:
any confidentiality concerns are vacuous if the communication partner is not genuine. In the context of MPC, not only compute nodes that handle the data, but also data providers and receivers need to be authenticated.
The former is required to increase trust in the input data and potentially achieve accountability, while the latter is needed to ensure that only eligible parties may request computations.

However, out-of-the-box authentication methods are not always applicable in certain scenarios, as the identity of data sources and data receivers may subject to data protection requirements. For example, it may be desired to determine only the eligibility to request a computation, but not the actual identity.
Yet, in case of misuse, methods for accountability may be needed.

The situation is further complicated when the data is managed on behalf of the owner by a third party (data custodian); when the owner is not able or willing to manage their own data. In this case, authentication would also be handled by the data custodian, with the owner first granting the right to do so.

To support large scale adoption, compatibility with governmental identities such as the upcoming European eIDAS 2.0 regulation is also necessary.

\paragraph{C3. Data usage policies.}
Precise data usage policies play a critical role in increasing trust and achieving acceptance by end users, particularly when personal or confidential data is involved.

Such policies describe the permissible ways in which data can be utilized, encompassing aspects such as eligible groups of receivers, temporal restrictions, requirements on the MPC or FHE setup (e.g., threshold, geographical distribution of nodes or preferred provider respective technology to use), the computation to be carried out (e.g., certain statistics including the required sample size or validation mechanisms), or data retention.

However, formulating and enforcing effective data usage policies presents several challenges.
These include striking a balance between maximizing data utility for innovation and safeguarding privacy rights, achieving high usability also for end users, addressing evolving technological advancements and data-sharing practices, and ensuring transparency and accountability.
Additionally, changing legal and market situations need to be addressable, potentially without re-involving data subjects in asynchronous scenarios.

\paragraph{C4. Node selection.}
The security of any MPC deployment crucially depends on the involved compute nodes, as well as the selected parameters (i.e., threshold and number of nodes).
Interestingly, although FHE is a completely different approach it shares many similarities when it comes to the requirements for the deployment, if more stakeholders are involved, e.g. decryption threshold and number of nodes for multi-party secret key generation.
However, the big difference is in the compute nodes needed to perform the actual computation.
For FHE the computation on the ciphertext can be done on a single server and without interaction, contrary to MPC which requires communication between compute nodes.
Nevertheless, the computational power needed on the compute nodes to run MPC protocols is less for communication intensive protocols and in general far less than for executing FHE.
Therefore, in the future hardware support is envisioned for FHE to accelerate computations and reduce power consumption, contrary to MPC, which does not require special purpose hardware.

Besides the general requirements on the node selection for MPC and FHE, also the time dependence plays a crucial role in the complexity of system management.
In certain (mainly static) scenarios, the selection of these nodes can be done once and (almost) forever.
However, the situation is very different in highly dynamic scenarios where data from many data sources is used as input, as each of them pose certain constraints on node selection.
Furthermore, compute nodes may be offered on an ``as-a-service'' basis by market players, such that their availability may have temporal variety.
Therefore, any mechanism for node selection needs to take these requirements into consideration.

In combination with the identity management challenges mentioned before, it further needs to be guaranteed that the involved nodes are not (potentially indirectly) controlled by a single legal entity.

This immediately also poses the question who decides, which nodes to involve.
If this process relies on a central entity, appropriate measures to minimize the required trust should be taken, e.g., by aiming for transparency of performed computations, or by having compute nodes verify usage policies without compromising privacy.
On the other hand, if this process is performed in a federated way, a circular argument (who chooses the participants of this set of entities) should be avoided.

\paragraph{C5. Access control.}
In static situations characterized by fixed computations and entities, it is often predetermined which inputs and outputs must be accessible to each party.
In this case, data providers may, e.g., encrypt input shares directly for designated compute nodes, which in turn encrypt the output for the specified data recipient.

Yet, in dynamic environments, this predictability may not always hold true.
Thus, if it is unknown upfront which (or how many) MPC nodes will execute a given computation -- and nodes might engage in computations on the same data across different sessions -- appropriate technologies must be implemented to ensure that the shares for these nodes can be derived as needed without compromising privacy.

A fundamental challenge lies in avoiding dependence on a single trusted entity or a single point of failure, necessitating careful design of key management procedures. Moreover, it is essential to guarantee that nodes cannot receive multiple consistent shares when the same input data is utilized in multiple computations involving the same node.

\paragraph{C6. Trustless intermediaries.}
Advancements in cryptographic techniques such as presented MPC and FHE methods are essential for enabling trustless data intermediaries.
These techniques ensure that data remains private and secure throughout the processing lifecycle, even in environments where trust cannot be assumed.

If integrated properly, data intermediaries can directly benefit for the secure computation capabilities in the data space and only store encrypted data locally.
However, for regulatory reasons, data intermediaries must meet several stringent requirements.
These include robust authentication and identity management systems to ensure that only authorized parties participate in computations, as well as precise definitions and enforcement mechanisms for data usage policies.
Furthermore, the selection of compute nodes must also be carefully managed accordingly to keep data encrypted over the whole life-cycle.

Thus, clear protocols must be established regarding who can access what data and with which keys.
This includes defining roles and permissions within the intermediary framework to ensure that access to both data and keys is tightly controlled and monitored.
The challenge of key management is especially pronounced in trustless environments, where no single entity is trusted with full control over the keys.
Innovative solutions, such as distributed key management systems, may be necessary to mitigate these risks, ensuring that no single point of failure can lead to a security breach.
Moreover, the access control systems must be designed to enforce these restrictions rigorously, allowing only authorized entities to perform decryption or initiate computations.
By meeting these requirements, data intermediaries can securely manage and process sensitive data, maintaining privacy and security in complex and distributed environments.

\section{Integration into Data Spaces}\label{sec:integration}\label{sec:mpc-integration}

\begin{figure}[t!]
	\includegraphics[width=\columnwidth]{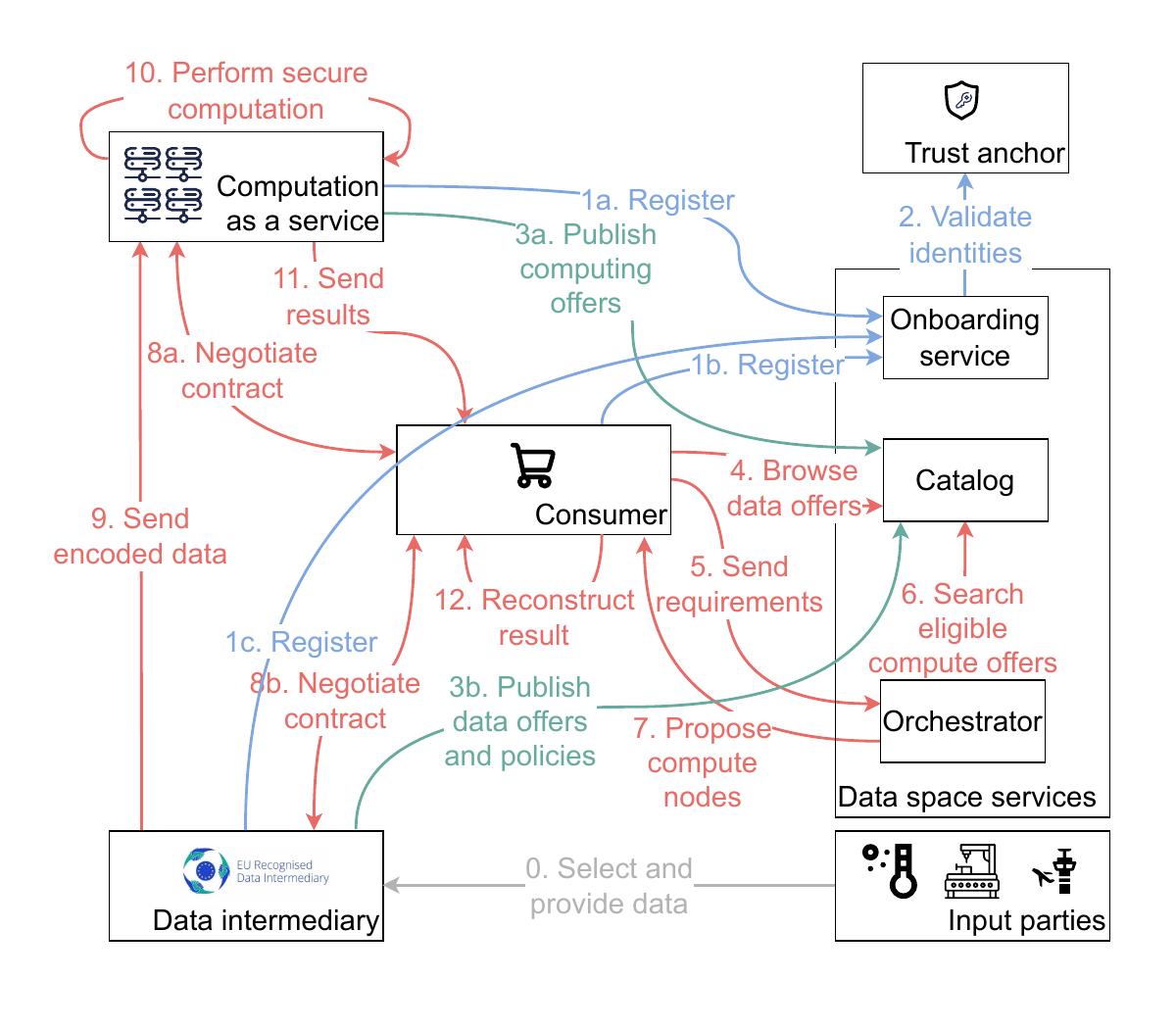}
	\caption{Components of the proposed data space-based deployment. \textcolor{blue}{\bf Blue}: onboarding phase; \textcolor{teal}{\bf green}: setup phase; \textcolor{red}{\bf red}: transaction phase. Based on initial ideas from \cite{siska_integrating_2024} but extended for additional stakeholders and technologies.}
	\label{fig:mpc_dataspace_architecture}
\end{figure}

We propose using data spaces as a basis to deploy secure multiparty computing in a dynamic scenario; that is, where some or all elements (stakeholders, input data, algorithm) are not known in advance.
Our goal is to create an ecosystem where participants can offer MPC-related assets under well-defined conditions ("policies"): input datasets, compute nodes or algorithms (intended function to be computed).
Other participants may consume these offers by running a computation on a chosen set of input datasets and compute nodes, while respecting the conditions set by the providers of these assets.
We divide the deployment of such a system in three phases: onboarding (participants), (asset) setup, and the transaction phase, where a single computation is executed.
The overall architecture is shown in \cref{fig:mpc_dataspace_architecture}.

\subsection{Onboarding and Setup Phase.}
First, participants need to be onboarded to the system (data space), which includes checking their identity and issuing some form of a proof of membership. At this phase, the identity of participants may be checked, possibly connecting to external trust anchors (TAs), see also challenge C2.

Second, onboarded participants may publish assets in the data space, potentially through the data intermediary representing them.
For the computation resource providers (i.e., for MPC or FHE), these include input data, compute nodes or even intended functions, each described by asset-specific metadata and associated with an individual policy that describes how they can be used, cf. C3.
Note that in a fully dynamic setting, both steps of the setup are also dynamic: participants and offers may be added, modified or removed during the lifetime of the data space.

\subsection{Transaction Phase.}
In the last phase, the actual transaction may occur.

\paragraph{Offer selection and contract negotiation.}
First, participants (potential consumers) may browse available offers and select a combination of input data, compute nodes and a function they would like to evaluate.
When selecting compute nodes, the consumer may pick offers explicitly or define conditions that nodes need to satisfy, depending on the selected computing paradigm.
For instance, for MPC, they might define that not all nodes should be hosted on the same server, or that all nodes have to be hosted in Europe.
From a usability point of view, it could also be desirable to offer some preconfigured choices relevant in different domains \cite{framner_making_2019} and from a performance standpoint also network or performance requirements could be included in the node selection, e.g., latency <30ms between nodes.
However, if the MPC nodes are not concretely defined, the orchestrator service may pick a random selection of compute nodes on offer, to satisfy such criteria.
Similar choices and configurations might be done for FHE-based services, e.g., regarding computation resources or location of the server to avoid the transfer of (encrypted) data, e.g., outside the EU.

In either case, the request is sent to data intermediary representing the owner of the respective offers as a contract request, after which an automatic contract negotiation process takes place to validate that all requirements with regards to the policies are met.
If this is the case, a contract between all parties is signed and the computation can be triggered.
Validation of conditions may happen via a service ("MPC/FHE orchestrator") offered by the data space authority, which can be the same service orchestrating MPC computation, cf. also C4.

\paragraph{Input provisioning.}
After all parties agreed to the transaction, the actual computation is started.
Therefore, the input data has to be read by the compute parties in encoded form.
Depending on the configuration, this step can be done either synchronously by the input parties sending the inputs to the compute nodes, but also asynchronously, if the data have been stored at a data custodian.
In this case, for security reasons and following the zero-trust principle, the data should only be stored in encrypted form.
However, this is not trivial, if the receiving compute nodes are not known in advance, cf. also C5.

Furthermore, to be more flexible, it is also desirable to delay the time of encoding if possible.
Thus, we distinguish immediate and late encoding.

\textit{Immediate encoding} is the naive way to generate input data by encoding the data prior to encrypting it for storage at the data custodian in the case of MPC.
Then each compute node only has to decrypt his received data fragment during input processing.
This is easier from a technological point of view, but less flexible and produces more overhead: as each share is encrypted individually, the total amount of data to be stored is large.
Additionally, MPC system parameters and encoding scheme have to be defined in advance.

In the case of FHE, immediate encoding means, that the secret key is already defined, and input can be already decrypted under the corresponding public key before storing it in the corresponding system.

In \textit{late encoding}, the plaintext is directly encrypted and stored at the data custodian, also for MPC.
This significantly reduces the storage overhead and increases flexibility, as MPC parameters are decided during the transaction phase and not the setup phase.
However, it is also technically more challenging, because some form of flexible threshold decryption is needed.
A compromise would be to symmetrically encrypt the input data and then only encrypt the key with a threshold method.
This would also save storage space but require the compute nodes to first decrypt the data obliviously \cite{lorunser_performance_2020}.

Similar questions arise for FHE based late decoding.
For FHE data could be encrypted under a user-controlled key and upon request, a re-encryption policy can be shared with the system which allows to transform the ciphertext to be used in a FHE computation under a dedicated key different to the user key.

\paragraph{Protocol execution.}
During computation the agreed MPC protocol is executed among the agreed nodes to compute the intended function on the data.
For FHE a single compute node is sufficient to compute the intended function, eventually by also receiving an additional evaluation key.
Although the step is rather straightforward, from a data space perspective it is important that the protocols available are standardized.
Policies can only be practically enforced, if wide interoperability among MPC/FHE nodes available in the ecosystem is guaranteed and enough stakeholders publish offers.
Additionally, to executing the MPC protocols or specific FHE schemes, plugins may also be of use.
If verifiability is a requirement, an additional zero-knowledge proof has to be generated by the system, posing additional challenges for policy definition, the capabilities of the MPC or FHE nodes, and the trustworthiness of required parameters, cf. also C1, C3, and C4.

Additionally, managing the leakage budget for secure computations which are intrinsic to the compute function by methods from differential privacy could also require for a plugin.

\paragraph{Post-processing.}
Finally, after the computation the results are held by the compute nodes in encrypted form.
To recover the plaintext output the ciphertexts have to be communicated (synchronously or asynchronously) to one or more result parties, which are allowed to learn the outcome of the computation.
For MPC the result is typicially reconstructed from fragments received from compute nodes, especially for secret sharing based protocols.
In the case of FHE, if a single result party was used and FHE computation was done under her respective keys, then she can directly decrypt the ciphertext from the compute node.
However, if threshold decryption was defined in the policy, the plaintext can only be decrypted by multiple parties together, which closely resembles the MPC setting.

Finally, after the reconstruction of the result succeeded additional post-computation validation, logging, and payment could take place to finalize the transaction.

In summary, by our comprehensive integration proposal of secure computing methods, i.e., MPC and FHE, into data spaces, we have shown the complexity we are facing when we go beyond the naive approach where dedicated parties with profound technology knowledge run a specific instance of a protocol.
However, this extra effort is necessary to make the system interoperable, being compatible with data spaces, and to leverage the MPC-as-a-service approach to lower the barriers for adoption.

\section{Technical Solutions}\label{sec:solutions}

In this section technical methods to solve the identified challenges are discussed.
We identify gaps in the state-of-the-art, present potential avenues to address the challenges and highlight where additional research is needed.


\subsection{Global Parameters}
To minimize the necessary trust, global parameters should be setup in a way that does not give any sufficiently small set of entities the possibility to control the choice of parameters.
Different approaches for this can be found in the literature.

One option are so-called \emph{setup ceremonies}, where a group of entities jointly generates parameters that are later needed for cryptographic protocols, thus ensuring the trustworthiness of the outcome.
Such ceremonies have been implemented for a variety of applications, including, e.g., the ZCash crypto currency\footnote{\url{https://zkproof.org/2021/06/30/setup-ceremonies/}}.

Another active research field in cryptography is focusing on so-called subversion resilience, where at least partial security guarantees can also be achieved if, e.g., a common reference string (CRS) cannot be trusted, e.g., \cite{DBLP:journals/joc/AbdolmalekiLSZ21}.

Other works, e.g., \cite{DBLP:conf/cans/BagheryS21}, consider the updatable CRS model, where users can update the CRS at any time, provided they demonstrate the correctness of the update.
The new CRS can then be deemed trustworthy (i.e., uncorrupted) as long as either the previous CRS or the updater was honest.
If multiple users partake in this process, it's possible to obtain a sequence of updates by different individuals over time.
If any update in the sequence is honest, the scheme remains sound.

\subsection{Authentication and Identity Management}
Authentication and identity management can differ between data spaces and may rely on traditional centralized (e.g. via a user database based on LDAP or Active Directory) or decentralized (e.g. using Decentralized Identifiers and Verifiable Credentials (VCs)) approaches. In any case, an onboarding process needs to be defined as part of data space governance, where the identity of participants is validated before granting them membership. The validation step normally relies on external trust anchors (e.g., eIDAS, DV SSL, GLEIF), with accepted trust anchors defined by the given data space's governance framework. As part of the onboarding process, participants may also record their public key and prove their control over it, providing a basis for a secure communication channel.

For instance, for Gaia-X \cite{gaia-x_framework_2023}, aspiring participants would submit their data as defined in the Trust Framework (e.g., ID, public key, address) to one of the Gaia-X Digital Clearing Houses (GXDCH) and receive a VC that they can use as proof. Internally, the GXDCH applies multiple validation checks, such as compatibility with the required metadata schema and validation via accepted trust anchors.


When a data custodian (ensuring data accessibility and security for a data owner) is also part of the system, the data owner first needs to authorize the custodian to act on their behalf. This can happen outside the data space context, via a separate contract between these parties, or is part of a data space service offering. The custodian then participates in the data space on behalf of the data owner. A more formalized and regulated instance of a data custodian is a Data Intermediary as defined in the Data Governance Act\footnote{\url{https://eur-lex.europa.eu/legal-content/EN/TXT/HTML/?uri=CELEX:32022R0868}} - while a data custodian focus on the technical and security aspects of data management the data intermediary facilitates data sharing and usage in compliance with legal and regulatory frameworks.

\medskip

While strong authentication may be required in many application cases, some scenarios require a delicate balance between privacy and authenticity, e.g., when an entity needs to fulfill a data usage policy but does not want to reveal its identity.
This can be achieved, e.g., using attribute-based credentials \cite{DBLP:conf/scn/CamenischL02,DBLP:conf/sacrypt/CamenischKLMNP15,DBLP:conf/eurocrypt/TessaroZ23a} letting parties prove statements about their attributes without revealing them in the plain.
In particular, this also covers selective disclosure as considered by W3C\footnote{\url{https://w3c-ccg.github.io/data-minimization/}} or EBSI\footnote{\url{https://ec.europa.eu/digital-building-blocks/sites/display/EBSI/Selective+Disclosure\%3A+An+EBSI+Improvement+Proposal}}.

Furthermore, somewhat similar to direct anonymous attestation (DAA)~\cite{DBLP:conf/ccs/BrickellCC04} or Intel's Enhanced Privacy ID (EPID)\footnote{\url{https://www.intel.com/content/www/us/en/developer/articles/technical/intel-enhanced-privacy-id-epid-security-technology.html}}, in order to increase reliability in data without compromising security, concepts like privacy-enhancing group signatures~\cite{DBLP:conf/IEEEares/KrennSS19,DBLP:conf/pkc/DiazL21} could be used. These let data sources such as sensors sign data to prove that it was generated using a genuine device, while keeping the precise identity of the device confidential.
MPC over authenticated inputs is also considered by~\cite{cryptoeprint:2022/1648}.

\subsection{Data usage policies}\label{subsec:policies}
Usage Control\cite{Jung2022} plays an important role in the enforcement of data policies, particularly in complex data environments. One of the significant challenges for MPC and FHE in data spaces is ensuring that data policies are effectively enforced throughout the data processing lifecycle. While the Open Digital Rights Language (ODRL)\footnote{\url{https://www.w3.org/TR/odrl-model/}} offers a flexible mechanism for defining permissions, prohibitions, and duties concerning digital content and services, its effectiveness is limited in the context of MPC where data processing involves complex computations across multiple data owners. The enforceability of these policies becomes even more complicated when considering the simpler, yet enforceable, nature of Rego\footnote{\url{https://www.openpolicyagent.org/docs/latest/}} within the Open Policy Agent (OPA) framework, which may not fully cater to the legal nuances required in MPC scenarios.

Moreover, the integration of secure computation as a service within data spaces necessitates a high degree of interoperability between different policy standards and legislative frameworks, no matter if MPC or FHE is considered.
The diverse landscape of standards like the Data Privacy Vocabulary (DPV)\footnote{\url{https://w3c.github.io/dpv/dpv/}} for expressing policies related to personal data processing, and international standards such as ISO/IEC 29184 and ISO/IEC 27560 for online privacy and data sharing, must be seamlessly aligned to support the complex operations of MPC and FHE respectively.

Compliance poses another challenge, especially with the introduction of legislative frameworks such as the Data Governance Act and the Data Act. These acts introduce new concepts like data intermediaries and data altruism, which, while enriching the data ecosystem, also add layers of complexity in ensuring that MPC services adhere to these regulations. Additionally, the empowerment of individuals through platforms like SOLID\footnote{\url{https://solidproject.org/}}, granting them control over their data, intersects with the operational dynamics of cryptographically protected computations, requiring robust mechanisms to ensure that user consent and data usage terms are respected in a multi-stakeholder environment.

Incorporating also the Data Catalog Vocabulary (DCAT)\footnote{\url{https://www.w3.org/TR/vocab-dcat-3/}} into the ecosystem of data spaces, to facilitate the discovery and interoperability of datasets, makes integrating usage polices even more challenging but also leads to a convergence of standards and practices for the participating stakeholders. By establishing a common framework, DCAT can serve as a tool in bridging the gap between different data policy standards. This convergence simplifies the process of managing and enforcing data usage policies across multiple platforms and jurisdictions, promoting a more unified and efficient approach to data sharing and processing.

\subsection{Node Selection}
In contrast to the permissionless systems prevalent in the blockchain world (e.g., Enigma, Partisia), data space services require registration, meaning they operate within a permissioned environment, thereby providing significant benefits with regards to node selection.

Nodes or node operators must be registered, and each node will be assigned with attributes describing its abilities.
Besides standards capabilities, like supported protocols, connection parameters like bandwidth, compute capabilities and other functional parameters, nodes must also be assigned with trust parameters.
Every node must be assigned to an identity, geo location, and trust zones, to enable automatic matching of compute task policies and nodes.
Despite the common attributes required for secure computation in general, we also have specific ones for the different technologies, MPC and FHE in our case.

For MPC we must support flexible definitions on the composition of a computing environment from multiple nodes which fulfil the non-collusion assumption best but still provide enough connectivity (network bandwidth and latency) to ensure efficient performance.
The following sample settings illustrate policies that shall be supported in an MPC-ready data space:
\begin{itemize}
	\item Nodes must be from 3 different entities in three different countries
	\item All nodes must be from the same country but from three different institutions or trust boundaries
	\item Nodes must have latency $\leq 10$ms but be from different trust zones
\end{itemize}
It is also of interest to combine basic attribute-based matching with random assignment capabilities for additional robustness.
Given the policy settings above, it should be possible to randomly assign nodes from all available combinations for different functions or even sub-functions, thereby also preventing sybil attacks.

In the case of FHE typically only a single node has to be agreed on to run the actual computation, except for the case of partitioning and parallelization of a task onto multiple nodes.
Here, hardware support could be essential and also support for more advanced methods for data input and output encoding.
However, when support for multi-key FHE or threshold decryption is needed, additional nodes need to be also defined to be part of the input encoding or decryption process which makes the configuration similarly challenging as for the MPC case.

In essence, node selection introduces many aspects which are interesting to support to make the infrastructure more trustworthy but also open and flexible.
It enables users to define their deployment requirements and enables participants to contribute compute resources to be used by others in a seamless way.

\subsection{Access Control}
An integral aspect of data usage policies is the delineation of authorized users' access to specific datasets.
While contractual enforcement suffices in numerous practical scenarios, there's a growing preference for technical solutions.
This approach, e.g., obviates the need for a data custodian to possess plaintext access to users' sensitive data.

In the following we sketch two options that realize this goal by leveraging advanced cryptographic methods beyond what was already discussed before.

One option following the late encoding approach could be to let data owners encrypt their data under their own public key using a so-called proxy re-encryption scheme \cite{DBLP:conf/eurocrypt/BlazeBS98,DBLP:conf/asiacrypt/ZhouLHZ23}. This allows the data custodian to transform ciphertexts under the user's public key into ciphertexts under a compute node's public key, provided that the user previously handed a so-called re-encryption key to the data custodian.
In case that the encryption scheme supports a homomorphic operation on ciphertexts consistent with the secret sharing scheme, the data custodian could now derive the shares for the selected compute nodes ad-hoc, without ever requiring to access the plaintext.
One drawback of this approach is, however, that the user has to derive individual re-encryption keys for all possible compute nodes, which may exclude nodes joining the ecosystem after the user making their data offer.

An alternative option based on early encoding leverages attribute-based encryption (ABE) \cite{DBLP:conf/eurocrypt/SahaiW05,DBLP:conf/eurocrypt/HohenbergerLWW23}.
In an ABE scheme, each participant receives a secret key linked to some attributes (e.g., geographical location), while ciphertexts are linked to policies.
A secret key can now only decrypt a ciphertext if the attributes of the secret key satisfy the policy of the ciphertext.
For instance, users could encrypt their shares according to their requirements (e.g., each share with a specific country);
while each compute node would receive a secret key linked to the country of its location.
Assuming proper identity management, doing so could cryptographically enforce that only compute nodes located in specific countries could decrypt certain shares, thereby enforcing that nodes from different legislations participate in a computation.
The main limitation of this approach is, that the master secret key, from which the individual secret keys are derived, needs to be administered securely and trustworthy within the MPCaaS ecosystem, e.g., by distributing it among several nodes which engage in an MPC protocol to derive novel keys for joining nodes.
Furthermore, the encoding scheme required for the computations need to be known in advance.

\subsection{Trustless Intermediaries}

A trustless data intermediary is a solution that facilitates data sharing and processing between parties without the need for mutual trust. By leveraging advanced cryptographic techniques such as Multi-Party Computation (MPC) and Fully Homomorphic Encryption (FHE), these intermediaries ensure data privacy and security even when the intermediary itself is not trusted by the involved parties. This approach is particularly valuable in scenarios where sensitive data must be processed or shared across organizational boundaries, and where traditional trust models are insufficient or undesirable.

\textit{Multi-Party Computation (MPC)} maintains the privacy of each party's data, as only the final result of the computation is revealed, with no individual data points being exposed. The security and effectiveness of MPC relies heavily on the trustworthiness and reliability of the selection of compute nodes. In a static setting, where the computational environment is predefined, this process is streamlined, enabling asynchronous processing and reducing complexity.

\textit{Fully Homomorphic Encryption (FHE)} ensures that data remains encrypted throughout the computation process, providing robust privacy protection even during processing. This technique is particularly advantageous in scenarios involving highly sensitive data, such as healthcare or financial services, where preventing data exposure at all stages is critical. FHE's capability to securely process data in both fixed and dynamic deployment scenarios makes it highly adaptable. In a static setting, where participants and computational environments are predefined, the secure management of evaluation keys becomes more straightforward, enabling trustless intermediaries to efficiently handle asynchronous data processing.

The challenge of key management is central to the successful deployment of trustless data intermediaries. It is essential to establish clear protocols that dictate who can access what data and which keys. In static deployment scenarios, the intermediary's ability to securely manage and access evaluation keys simplifies asynchronous processing and enhances overall system security. This controlled environment ensures that sensitive data is processed and shared without compromising privacy or security.

In conclusion, trustless intermediaries using cryptographic techniques such as MPC and FHE, offer a robust solution for secure data sharing and processing in contexts where traditional trust models fall short. By focusing on the secure management of nodes / keys and considering the static or dynamic nature of the deployment, organizations can effectively leverage these technologies to protect sensitive data while enabling valuable data-driven collaboration.

\section{Conclusion}\label{sec:conclusion}

This paper presents a comprehensive approach for integrating secure multiparty computation (MPC) and fully homomorphic encryption (FHE) into data spaces, laying the groundwork for secure and trustworthy data sharing in the future Data Economy.
The authors address various challenges and their potential solutions, namely global parameters, authentication and identity management, data usage policies, node selection, trustless data intermediaries, and access control.
By adopting these solutions, organizations can enhance privacy and security while facilitating data sharing in dynamic environments.

Moreover, we also discuss the impact of data intermediaries, which will play an important role in this framework by bridging the gap between data providers and consumers.
They enable secure data processing even in scenarios where trust is minimal or non-existent.
Through the application of MPC and FHE, these intermediaries ensure that sensitive data can be shared and processed without compromising privacy or security, making them indispensable in cross-organizational collaborations.
Their function is vital for maintaining compliance with regulatory frameworks and ensuring that data transactions are both secure and efficient, thus fostering innovation and collaboration within the data economy.

However, several research gaps remain.
There is a pressing need for more efficient and scalable MPC protocols that can handle large-scale datasets effectively.
For FHE, protocol hardware support will be needed to support practically relevant performance for computation, however, this is on the horizon.
Additionally, dynamic and flexible access control mechanisms in distributed environments are essential to address the evolving needs of data usage. Privacy concerns related to potential information leakage during protocol execution also require further exploration.
Moreover, the development of standardized and interoperable frameworks will be critical to support MPC- and FHE-enabled data spaces across various domains and applications.
By overcoming these challenges and fully leveraging the capabilities of data intermediaries, the potential for secure, privacy-preserving data sharing in the Data Economy can be realized.
Further research and development efforts are needed to overcome these challenges and ensure the successful adoption of this approach in practice.

%

\section*{Acknowledgments}

This work was in part funded by the European Union under the HORIZON SESAR JU Grant Agreement No. 101114675 ({\sc HARMONIC}),
where UK participants received funding from UK Research and Innovation (UKRI) under funding guarantee grant No. 10091990, and swiss partner from the Swiss State Secretariat for Education, Research and Innovation (SERI).
Additionally, it was supported by the Austrian Research Promotion Agency FFG within the {\sc Present} project.
Views and opinions expressed are however those of the author(s) only and do not necessarily reflect those of the funding agencies.
Neither the European Union, FFG, nor the granting authority can be held responsible for them.

\bibliography{bibliography/zotero.bib,bibliography/lore_mpcaas.bib,bibliography/dataspace.bib,bibliography/fhe.bib}

\end{document}